\documentclass[preprint,eqsecnum,aps,nofootinbib]{revtex4}
\usepackage[dvips]{graphicx}
\usepackage{amssymb}
\usepackage{amsmath}
\usepackage{amsmath, amsthm, amssymb, mathrsfs}
\usepackage{pstricks}

\begin{document}

\title{GHZ versus W : Quantum Teleportation through Noisy Channels}
\author{
Eylee Jung$^{1}$, Mi-Ra Hwang$^{1}$, You Hwan Ju$^{1}$, Min-Soo Kim$^{2}$, 
Sahng-Kyoon Yoo$^{3}$, Hungsoo Kim$^{4}$, 
D. K. Park$^{1}$, Jin-Woo Son$^{2}$, S. Tamaryan$^{5}$, 
Seong-Keuck Cha$^{6}$}

\vspace{1.0cm}

\affiliation{
$^1$ Department of Physics, Kyungnam University, Masan, 631-701, 
Korea  \\
$^2$ Department of Mathematics, Kyungnam University, Masan,
631-701, Korea         \\
$^3$ Green University, Hamyang, 676-872, Korea \\
$^4$ The Institute of Basic Science, Kyungnam University, 
Masan, 631-701, Korea \\
$^5$ Theory Department, Yerevan Physics Institute,
Yerevan-36, 375036, Armenia   \\
$^6$ Department of Chemistry, Kyungnam University, Masan, 631-701, Korea}

\begin{abstract}
Which state does lose less quantum information between GHZ and W states when they are
prepared for two-party quantum teleportation through noisy channel? We address
this issue by solving analytically a  master equation in the Lindbald form 
with introducing the 
noisy channels which makes the quantum channels to be mixed states. It is found that 
the answer of the question is dependent on the type of the noisy channel. If, for 
example, the noisy channel is ($L_{2,x}$, $L_{3,x}$, $L_{4,x}$)-type where 
$L's$ denote the Lindbald operators, GHZ state is 
always more robust than W state, i.e. GHZ state preserves more quantum information. 
In, however, ($L_{2,y}$, $L_{3,y}$, $L_{4,y}$)-type channel the situation becomes
completely reversed. In ($L_{2,z}$, $L_{3,z}$, $L_{4,z}$)-type channel W state is 
more robust than GHZ state when the noisy parameter ($\kappa$) is comparatively 
small while GHZ state becomes more robust when $\kappa$ is large. In isotropic 
noisy channel we found that both states preserve equal amount of quantum information. 
A relation between the average fidelity and entanglement for the mixed state quantum 
channels are discussed.
\end{abstract}


\maketitle

\section{Introduction}

Quantum teleportation is an important process in quantum information
theories\cite{nielsen00}. This process enables us to transmit an unknown 
quantum state from a sender, called Alice, to a remote recipient, called Bob, via
dual classical channels. Bennett et al have shown this process firstly in
Ref.\cite{bennett93} about fifteen years ago. In this paper authors used an
Einstein-Podolsky-Rosen state 
\begin{equation}
\label{EPR1}
|EPR\rangle = \frac{1}{\sqrt{2}} (|00\rangle + |11\rangle)
\end{equation}
as an quantum channel between Alice and Bob. In fact, $|EPR\rangle$ is not the only
two-qubit state which allows a perfect quantum teleportation. Any states that are
local-unitary(LU) equivalent with  $|EPR\rangle$ also can be used as quantum 
channels for the perfect teleportation. This set of states forms a set of maximally 
entangled states. 

Subsequently, quantum teleportation using three-qubit quantum channels are discussed.
In three-qubit system it is well-known that there are two LU-inequivalent 
types of the maximally entangled states, called 
Greenberger-Horne-Zeilinger(GHZ)\cite{green89} and W\cite{dur00-1} states whose 
general expressions are 
\begin{eqnarray}
\label{ghzw1}
& &|GHZ\rangle = a |000\rangle + b |111\rangle
\hspace{1.0cm} (|a|^2 + |b|^2 = 1)
\\   \nonumber
& &|W\rangle = a |001\rangle + b |010\rangle + c |100\rangle
\hspace{1.0cm} (|a|^2 + |b|^2 + |c|^2 = 1).
\end{eqnarray}
The perfect two-party quantum teleportation with exact GHZ state\footnote{Exact GHZ 
state is $|GHZ\rangle$ in Eq.(\ref{ghzw1}) with $a=b=1/\sqrt{2}$.} was discussed in 
Ref.\cite{karl98}. Furthermore, authors in Ref.\cite{karl98} discussed three-party
teleportation (Alice, Bob, Cliff) with the GHZ state. This can be used as an imperfect
quantum cloning machine\cite{wootters82}.

Recently, it was shown\cite{agra06-1} that not only GHZ state
\begin{equation}
\label{ghz1}
|\psi_{GHZ}\rangle = \frac{1}{\sqrt{2}} \left( |000\rangle + |111\rangle \right)
\end{equation}
but also W state
\begin{equation}
\label{W1}
|\psi_{W}\rangle = \frac{1}{2} \left( |100\rangle + |010\rangle + \sqrt{2} |001\rangle
\right)
\end{equation}
can be used as quantum channels for the perfect two-party teleportation.
As shown in Ref.\cite{tamaryan07-1} both $|\psi_{GHZ}\rangle$ and $|\psi_{W}\rangle$
have $G(\psi) = 1 / \sqrt{2}$, where $G(\psi)$ is a Groverian entanglement 
measure\cite{biham01-1}. Motivated from the fact that $|\psi_{GHZ}\rangle$ 
and $|\psi_{W}\rangle$ have same Groverian entanglement measure, authors in 
Ref.\cite{jung07-2} have shown that the state
\begin{equation}
\label{ours}
|\tilde{\psi}\rangle = \frac{1}{\sqrt{2}} \left( |00q_1\rangle + |11q_2\rangle \right)
\end{equation}
where $|q_1\rangle$ and $|q_2\rangle$ are arbitrary normalized one-qubit states, has
also $G(\psi) = 1 / \sqrt{2}$ and it can be used as a perfect two-party teleportation.

The fact that both $|\psi_{GHZ}\rangle$ and $|\psi_{W}\rangle$ allow the perfect 
two-party teleportation naturally arises the following question: which state is better
if noisy channels are introduced in the process of teleportation? The purpose of 
this paper is to address this issue by solving analytically a master equation in the 
Lindbald form\cite{lindbald76}
\begin{equation}
\label{lindbald}
\frac{\partial \rho}{\partial t} = -i [H_S, \rho] + \sum_{i, \alpha}
\left(L_{i,\alpha} \rho L_{i,\alpha}^{\dagger} - \frac{1}{2}
\left\{ L_{i,\alpha}^{\dagger} L_{i,\alpha}, \rho \right\} \right)
\end{equation}
where the Lindbald operator 
$L_{i,\alpha} \equiv \sqrt{\kappa_{i,\alpha}} \sigma^{(i)}_{\alpha}$ acts on the 
$i$th qubit and describes decoherence, where $\sigma^{(i)}_{\alpha}$ denotes the 
Pauli matrix of the $i$th qubit with $\alpha = x,y,z$. The constant $\kappa_{i,\alpha}$
is approximately equals to the inverse of decoherence time. The master equation
approach is shown to be equivalent to the usual quantum operation approach 
for the description of open quantum system\cite{nielsen00}.

\begin{figure}
\begin{center}
\includegraphics[height=5cm]{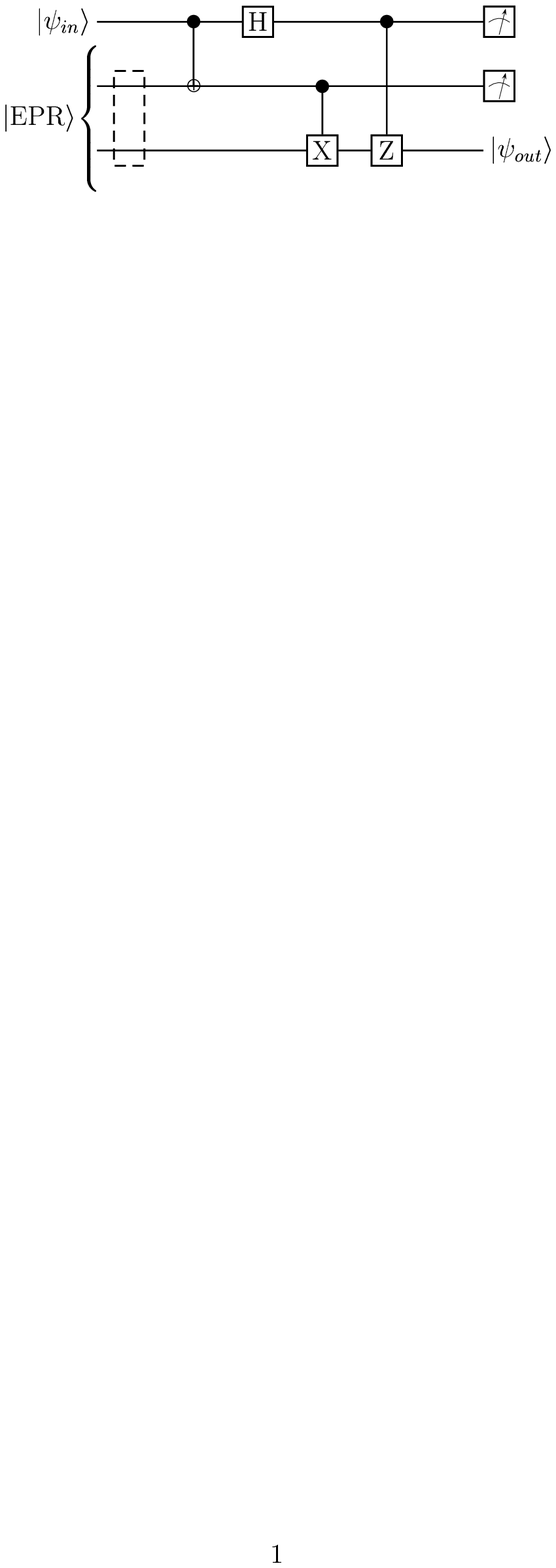}
\end{center}
\caption{A quantum circuit for quantum teleportation through noisy channels with
EPR state. The top two lines belong to Alice while the bottom line belongs to Bob.
The dotted box represents noisy channels, which makes the quantum channel to be mixed
state.}
\end{figure}

To reduce the effect of the noisy channels in the teleportation process the special 
purification protocols have been developed in Ref.\cite{bennett95,cheong07}. Via this 
purification process for the noisy quantum channel one can increase the fidelity of 
teleportation. One can also directly compute the fidelity between initial unknown state 
and final state. 
This was discussed in 
Ref.\cite{oh02} when two-qubit EPR quantum channel interacts with various noisy channels. 
The quantum circuit for teleportation with $|EPR\rangle$
through noisy channel is illustrated in Fig. 1. The two top lines belong to Alice, 
while the bottom one belongs to Bob. The dotted box denotes noisy channel. Although 
different noisy channels were discussed in Ref.\cite{oh02}, we will concentrate on 
the noisy channel which makes the quantum channel to be mixed because our main 
purpose is comparison of $|\psi_{GHZ}\rangle$ with $|\psi_W\rangle$ in the teleportation 
process.

How much quantum information is lost due to noisy channel can be measured by fidelity
between $|\psi_{in}\rangle$ and $|\psi_{out}\rangle$. In order to quantify this 
quantity it is more convenient to use the density matrix. Let 
$\rho_{in} = |\psi_{in}\rangle \langle \psi_{in}|$ and 
$\rho_{EPR} = |EPR\rangle \langle EPR|$. Then the density matrix for output state
reduces to 
\begin{equation}
\label{out1}
\rho_{out} = \mbox{Tr}_{1,2} \left[ U_{EPR} \rho_{in} \otimes \varepsilon (\rho_{EPR})
U_{EPR}^{\dagger}
\right]
\end{equation}
where $\mbox{Tr}_{1,2}$ is partial trace over Alice's qubits and $U_{EPR}$ is an
unitary operator implemented by quantum circuit in Fig. 1. In Eq.(\ref{out1}) 
$\varepsilon$ denotes a quantum operation which maps from $\rho_{EPR}$ to 
$\varepsilon (\rho_{EPR})$ due to noisy channel. The explicit expressions for 
$\varepsilon (\rho_{EPR})$ can be derived by solving the master equation.
Then the quantity which measures how much information is preserved or lost can be 
written as 
\begin{equation}
\label{fidelity1}
F = \langle \psi_{in} | \rho_{out} | \psi_{in}\rangle
\end{equation}
which is the square of the usual fidelity 
$F(\rho, \sigma) = \mbox{Tr} \sqrt{\rho^{1/2} \sigma \rho^{1/2}}$.
Thus $F=1$ implies the perfect teleportation. If $1 - F$ becomes larger and larger,
this indicates that we lost quantum information more and more.

The paper is organized as follows. In section II we consider the two-party quantum
teleportation with $|\psi_{GHZ}\rangle$ as quantum channel when noisy channel makes 
$|\psi_{GHZ}\rangle$ to be mixed state. Solving the master equation (\ref{lindbald})
analytically, we compute $F$ in Eq.(\ref{fidelity1}) explicitly when Lindbald 
operators are
($L_{2,x}$, $L_{3,x}$, $L_{4,x}$), ($L_{2,y}$, $L_{3,y}$, $L_{4,y}$),
($L_{2,z}$, $L_{3,z}$, $L_{4,z}$) and isotropy respectively. In section III 
calculation in previous section is repeated with changing the quantum channel from
$|\psi_{GHZ}\rangle$ to $|\psi_W\rangle$. In section IV the results of section II 
and section III are compared with each other. It is shown that the answer of the 
question  
`` which state is more robust\footnote{Throughout this paper ``a given state is more 
robust'' means that the state does lose less quantum information in the quantum 
teleportation through noise channels} in the noisy channel?'' 
is completely dependent on the type
of the noisy channel. In ($L_{2,x}$, $L_{3,x}$, $L_{4,x}$), for example, 
$|\psi_{GHZ}\rangle$ preserves more information than $|\psi_{W}\rangle$ while reverse
situation occurs in ($L_{2,y}$, $L_{3,y}$, $L_{4,y}$) channel. The situation in 
($L_{2,z}$, $L_{3,z}$, $L_{4,z}$) channel is more delicate. When the multiplication of 
noisy parameter
with time parameter, i.e. $\kappa_{i,z} t$, is small, $|\psi_{W}\rangle$ is slightly more
robust than $|\psi_{GHZ}\rangle$. If, however, $\kappa_{i,z} t$ becomes larger, 
$|\psi_{GHZ}\rangle$ preserves more information than $|\psi_{W}\rangle$. In isotropy
noisy with $\kappa_{i,x} = \kappa_{i,y} = \kappa_{i,z} = \kappa$ the average of 
$F$ over all input state $|\psi_{in}\rangle$ becomes identical for $|\psi_{GHZ}\rangle$
and $|\psi_{W}\rangle$. In section IV we give a brief conclusion. Also we discuss 
in this section a relation between average fidelity and entanglement for the mixed state
quantum channels.


\section{GHZ state with noisy channels}

\begin{figure}
\begin{center}
\includegraphics[height=5.0cm]{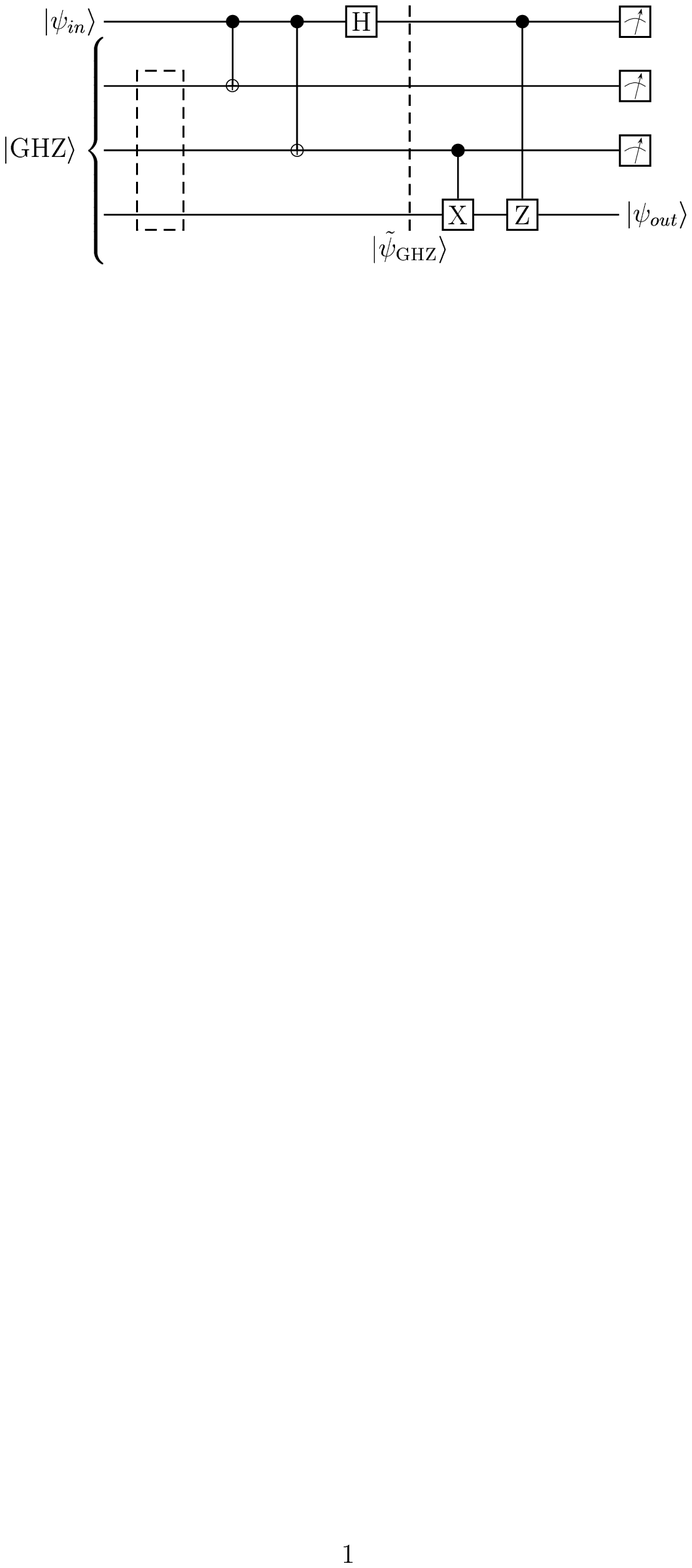}
\end{center}
\caption{A quantum circuit for quantum teleportation through noisy channels with
GHZ state. The top three lines belong to Alice while the bottom line belongs to Bob.
The dotted box represents noisy channels, which makes the quantum channel to be mixed
state.}
\end{figure}


In this section we would like to explore the effect of the noisy channels when the 
teleportation is performed with $|\psi_{GHZ}\rangle$. It is convenient to write the 
unknown state $|\psi_{in}\rangle$ to be teleported as a Bloch vector on a Bloch 
sphere in a form:
\begin{equation}
\label{unknown}
|\psi_{in}\rangle = \cos \left(\frac{\theta}{2}\right) e^{i \phi / 2} |0\rangle +
\sin  \left(\frac{\theta}{2}\right) e^{-i \phi / 2} |1\rangle
\end{equation}
where $\theta$ and $\phi$ are the polar and azimuthal angles.

The quantum circuit for teleportation with $|\psi_{GHZ}\rangle$ is shown in Fig. 2. 
The three top lines belong to Alice, while the bottom one belongs to Bob. The dotted
box denotes noisy channel. Comparing Fig. 2 to Fig. 1 there appears one more 
control-Not gate between the unknown state and GHZ state. 

The density for output state can be computed by 
\begin{equation}
\label{output-ghz1}
\rho_{out} = \mbox{Tr}_{1,2,3} \left[ U_{GHZ} \rho_{in} \otimes 
\varepsilon(\rho_{GHZ}) U_{GHZ}^{\dagger} \right]
\end{equation}
where $\mbox{Tr}_{1,2,3}$ is partial trace over Alice's qubits
and $U_{GHZ}$ is an unitary operator, which can be read straightly from Fig. 2.
In Eq.(\ref{output-ghz1}) $\rho_{in} = |\psi_{in}\rangle \langle \psi_{in}|$ and 
$\rho_{GHZ} = |\psi_{GHZ}\rangle \langle \psi_{GHZ}|$.

Now, we consider $(L_{2,z}, L_{3,z}, L_{4,z})$ noise channel because it is most 
simple to solve the master equation (\ref{lindbald}). Putting 
$\sigma_{ij} = \varepsilon_{ij} (\rho_{GHZ})$ with $i,j = 0, \cdots, 7$ and 
assuming $H_S = 0$ and $\kappa_{2,z} = \kappa_{3,z} = \kappa_{4,z} = \kappa$, the 
master equation reduces to $8$ diagonal and $28$ off-diagonal first-order differential
equations. Most of them simply give trivial solution and the only non-vanishing 
components are $\sigma_{00} = \sigma_{77} = 1/2$ and 
$\sigma_{07} = \sigma_{70} = e^{-6 \kappa t} / 2$. Thus in this noisy channel 
$\varepsilon(\rho_{GHZ})$ becomes 
\begin{equation}
\label{qoghz1}
\varepsilon(\rho_{GHZ}) = \frac{1}{2} \left( |000\rangle \langle 000 | + 
|111\rangle \langle 111| \right) + 
\frac{1}{2} e^{-6 \kappa t} \left(|000\rangle \langle 111| + |111\rangle \langle 000 |
\right).
\end{equation}
Inserting Eq.(\ref{qoghz1}) into Eq.(\ref{output-ghz1}) it is straightforward to 
derive $\rho_{out}$. Then the quantity $F$ defined in Eq.(\ref{fidelity1}) is dependent
on input angle $\theta$ as follows:
\begin{equation}
\label{f-ghz1}
F(\theta, \phi) = 1 - \frac{1}{2} \left(1 - e^{-6 \kappa t}\right) \sin^2 \theta.
\end{equation}
The average $F(\theta, \phi)$ over all possible input unknown states defined 
\begin{equation}
\label{average}
\bar{F} \equiv \frac{1}{4 \pi} \int_{0}^{2 \pi} d \phi \int_{0}^{\pi} d \theta 
\sin \theta F(\theta, \phi)
\end{equation}
reduces to 
\begin{equation}
\label{f-ghz2}
\bar{F} = \frac{2}{3} + \frac{1}{3} e^{-6 \kappa t}.
\end{equation}
It is easy to check that $F(\theta, \phi) = \bar{F} = 1$ when there is no noisy 
channel, i.e. $\kappa = 0$, which implies the perfect teleportation.

Next, we consider $(L_{2,x}, L_{3,x}, L_{4,x})$ noisy channel. Putting again
$\sigma_{ij} = \varepsilon_{ij} (\rho_{GHZ})$ and assuming again 
$H_S = 0$ and $\kappa_{2,x} = \kappa_{3,x} = \kappa_{4,x} = \kappa$, one can show that
the master equation (\ref{lindbald}) reduces to $8$ diagonal coupled linear 
differential equations and $28$ off-diagonal coupled linear differential equations. 
The $8$ diagonal equations imply 
$\sum_{i=0}^{3} \sigma_{ii} = \sum_{i=4}^{7} \sigma_{ii} = 1/2$. Thus we can write
$\sigma_{00} = 1/2 + \delta q_0$, $\sigma_{11} = \delta q_1$, 
$\sigma_{22} = \delta q_2$, $\sigma_{33} = -\delta q_0 - \delta q_1 - \delta q_2$,
$\sigma_{44} = \delta q_4$, $\sigma_{55} = \delta q_5$, $\sigma_{66} = \delta q_6$,
and $\sigma_{77} = 1/2 - (\delta q_4 + \delta q_5 + \delta q_6)$ with 
$\delta q_i (t=0) = 0$ for all $i$. Inserting these expressions into the original 
coupled equations, one can easily derive the diagonal components of $\sigma$, which
are 
\begin{eqnarray}
\label{sghz1}
& &\sigma_{00} = \sigma_{77} = \frac{1}{8} \left(1 + 3 e^{-4 \kappa t} \right)
\\  \nonumber
& &\sigma_{11} = \cdots = \sigma_{66} = \frac{1}{8} \left(1 - e^{-4 \kappa t} \right).
\end{eqnarray}

The off-diagonal $28$ coupled equations consist of $7$ groups, each of which are 
$4$ coupled differential equations in the closed form. Thus we can solve all of them
by similar way. Most of them give the trivial solutions and the nonvanishing 
components are 
\begin{eqnarray}
\label{sghz2}
& &\sigma_{07} = \frac{1}{8} \left(1 + 3 e^{-4 \kappa t} \right)
\\  \nonumber
& &\sigma_{16} = \sigma_{25} = \sigma_{34} 
= \frac{1}{8} \left(1 - e^{-4 \kappa t} \right)
\end{eqnarray}
with $\sigma_{ij} = \sigma_{ji}$. 
Defining 
\begin{eqnarray}
\label{def-ghz1}
& &\alpha_{+} \equiv 1 + 3 e^{-4 \kappa t}  \\  \nonumber
& &\alpha_{-} \equiv 1 - e^{-4 \kappa t},
\end{eqnarray}
we can express $\varepsilon(\rho_{GHZ})$ analytically in a form:
\begin{eqnarray}
\label{qo-ghz2}
\varepsilon(\rho_{GHZ}) = \frac{1}{8}
\left(          \begin{array}{cccccccc}
\alpha_+ & 0 & 0 & 0 & 0 & 0 & 0 & \alpha_+   \\
0 & \alpha_- & 0 & 0 & 0 & 0 & \alpha_- & 0   \\
0 & 0 & \alpha_- & 0 & 0 & \alpha_- & 0 & 0   \\
0 & 0 & 0 & \alpha_- & \alpha_- & 0 & 0 & 0   \\
0 & 0 & 0 & \alpha_- & \alpha_- & 0 & 0 & 0   \\
0 & 0 & \alpha_- & 0 & 0 & \alpha_- & 0 & 0   \\
0 & \alpha_- & 0 & 0 & 0 & 0 & \alpha_- & 0   \\
\alpha_+ & 0 & 0 & 0 & 0 & 0 & 0 & \alpha_+ 
\end{array}
\right).
\end{eqnarray}
Then using Eq.(\ref{output-ghz1}) one can compute $F(\theta, \phi)$ and $\bar{F}$,
whose expressions are
\begin{eqnarray}
\label{f-ghz3}
& &F(\theta, \phi) = \frac{1}{2} \left[(1 + \sin^2 \theta \cos^2 \phi) + 
e^{-4 \kappa t} (\cos^2 \theta + \sin^2 \theta \sin^2 \phi)
\right]  
\\   \nonumber
& &\bar{F} = \frac{2}{3} + \frac{1}{3} e^{-4 \kappa t}.
\end{eqnarray}

Similar calculation shows that $\varepsilon(\rho_{GHZ})$ for 
$(L_{2,y},L_{3,y},L_{4,y})$ noisy channel becomes
\begin{eqnarray}
\label{qo-ghz3}
\varepsilon(\rho_{GHZ}) = \frac{1}{8}
\left(          \begin{array}{cccccccc}
\alpha_+ & 0 & 0 & 0 & 0 & 0 & 0 & \beta_1   \\
0 & \alpha_- & 0 & 0 & 0 & 0 & -\beta_2 & 0   \\
0 & 0 & \alpha_- & 0 & 0 & -\beta_2 & 0 & 0   \\
0 & 0 & 0 & \alpha_- & -\beta_2 & 0 & 0 & 0   \\
0 & 0 & 0 & -\beta_2 & \alpha_- & 0 & 0 & 0   \\
0 & 0 & -\beta_2 & 0 & 0 & \alpha_- & 0 & 0   \\
0 & -\beta_2 & 0 & 0 & 0 & 0 & \alpha_- & 0   \\
\beta_1 & 0 & 0 & 0 & 0 & 0 & 0 & \alpha_+ 
\end{array}
\right).
\end{eqnarray}
where $\alpha_{\pm}$ are given in Eq.(\ref{def-ghz1}) and, $\beta_1$ and $\beta_2$ are
defined as
\begin{eqnarray}
\label{def-ghz2}
& &\beta_1 = 3 e^{-2 \kappa t} + e^{-6 \kappa t}     \\   \nonumber
& &\beta_2 = e^{-2 \kappa t} - e^{-6 \kappa t}.
\end{eqnarray}
One may wonder why the off-diagonal components of Eq.(\ref{qo-ghz3}) is much different 
from those of Eq.(\ref{qo-ghz2}) because of the following consideration: 
if $\sigma_{ij}^x$ and $\sigma_{ij}^y$ are density matrices for $(L_{2x}, L_{3x}, L_{4x})$ and 
$(L_{2y}, L_{3y}, L_{4y})$ noises respectively, then 
$(u \otimes u \otimes u) \sigma_{ij}^x (u \otimes u \otimes u)^{\dagger}$ satisfies the master
equation for the $(L_{2y}, L_{3y}, L_{4y})$ provided that $u$ is an unitary operator 
satisfying $u \sigma_x u^{\dagger} = \sigma_y$. Although this is completely correct, this
does not guarantee 
$\sigma_{ij}^y = (u \otimes u \otimes u) \sigma_{ij}^x (u \otimes u \otimes u)^{\dagger}$ 
because 
$\sigma_{ij}^x$ and $\sigma_{ij}^y$ should satisfy the boundary condition, {\it i.e.}
$\sigma_{ij}^x = \sigma_{ij}^y = \rho_{GHZ}$ when $\kappa t = 0$. 
The detailed computation for the 
off-diagonal components of $\sigma_{ij}^x$ and $\sigma_{ij}^y$ is briefly summarized 
in appendix.

One can show that Eq.(\ref{output-ghz1}) generates
\begin{eqnarray}
\label{f-ghz4}
& &F(\theta, \phi) = \frac{1}{2} \left[1 + \sin^2 \theta \sin^2 \phi e^{-2 \kappa t}
+ \cos^2 \theta e^{-4 \kappa t} + \sin^2 \theta \cos^2 \phi
e^{-6 \kappa t} \right]
\\   \nonumber
& &\bar{F} = \frac{1}{6} \left(3 + e^{-2 \kappa t} + e^{-4 \kappa t} + e^{-6 \kappa t}
\right).
\end{eqnarray}

For isotropic noise, which is described by nine Lindbald operators, $L_{2,\alpha}$,
$L_{3,\alpha}$, and $L_{4,\alpha}$ with $\alpha = x,y,z$, $\varepsilon(\rho_{GHZ})$
becomes
\begin{eqnarray}
\label{qo-ghz4}
\varepsilon(\rho_{GHZ}) = \frac{1}{8}
\left(          \begin{array}{cccccccc}
\tilde{\alpha}_+ & 0 & 0 & 0 & 0 & 0 & 0 & \gamma   \\
0 & \tilde{\alpha}_- & 0 & 0 & 0 & 0 & 0 & 0   \\
0 & 0 & \tilde{\alpha}_- & 0 & 0 & 0 & 0 & 0   \\
0 & 0 & 0 & \tilde{\alpha}_- & 0 & 0 & 0 & 0   \\
0 & 0 & 0 & 0 & \tilde{\alpha}_- & 0 & 0 & 0   \\
0 & 0 & 0 & 0 & 0 & \tilde{\alpha}_- & 0 & 0   \\
0 & 0 & 0 & 0 & 0 & 0 & \tilde{\alpha}_- & 0   \\
\gamma & 0 & 0 & 0 & 0 & 0 & 0 & \tilde{\alpha}_+  
\end{array}
\right).
\end{eqnarray}
where
\begin{eqnarray}
\label{def-ghz3}
& &\tilde{\alpha}_+ = 1 + 3 e^{-8 \kappa t}   \\  \nonumber
& &\tilde{\alpha}_- = 1 - e^{-8 \kappa t}    \\  \nonumber
& &\gamma = 4 e^{-12 \kappa t}.
\end{eqnarray}
In this case $F(\theta, \phi)$ and $\bar{F}$ becomes
\begin{eqnarray}
& &F(\theta, \phi) = \frac{1}{2} \left[1 + e^{-8 \kappa t} \cos^2 \theta + 
e^{-12 \kappa t} \sin^2 \theta \right]   \\  \nonumber
& &\bar{F} = \frac{1}{6} \left(3 + e^{-8 \kappa t} + 2 e^{-12 \kappa t} \right).
\end{eqnarray}
It is interesting to note that $F(\theta, \phi)$ for the isotropic noisy channel is
dependent on angle parameter $\theta$, while same quantity is independent of $\theta$
in Ref.\cite{oh02}, where two-qubit EPR state was used. The final results of 
$F(\theta, \phi)$ and $\bar{F}$ are summarized at Table I and will be compared to those
derived from $|\psi_{W}\rangle$. In the next section we will discuss the effect of noisy 
channels when we prepare $|\psi_{W}\rangle$ for the quantum teleportation.


\section{W state with noisy channels}

\begin{figure}
\begin{center}
\includegraphics[height=5.0cm]{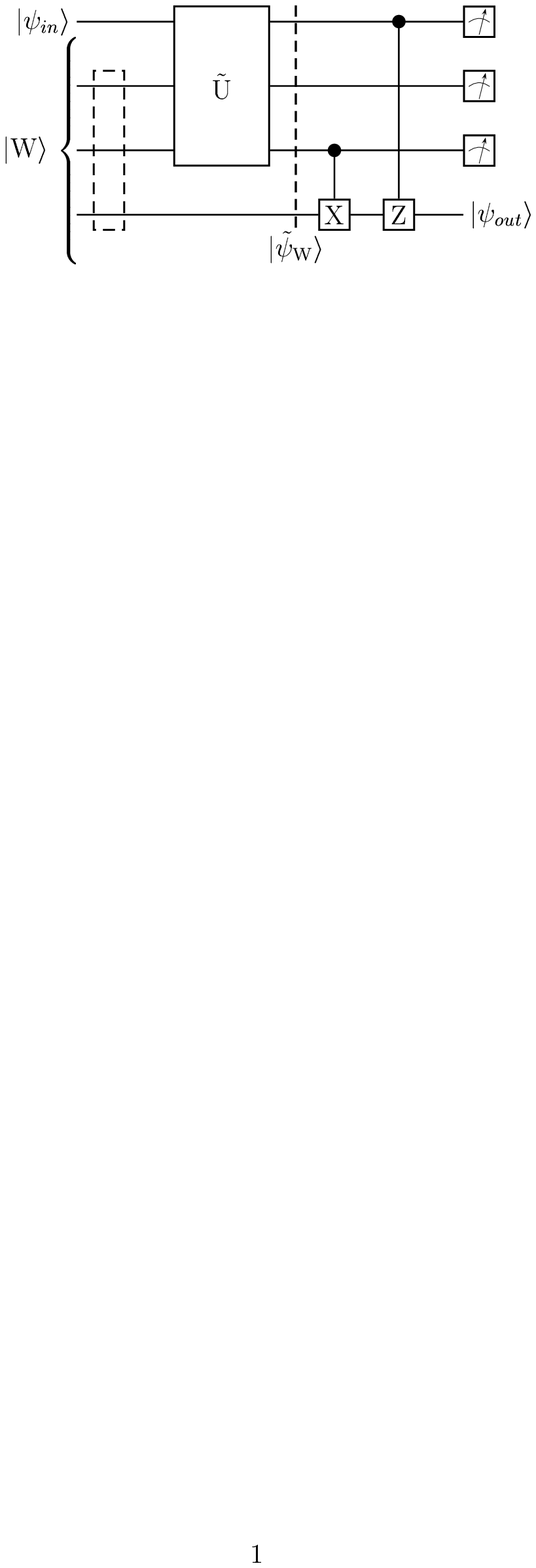}
\end{center}
\caption{A quantum circuit for quantum teleportation through noisy channels with
W state. The top three lines belong to Alice while the bottom line belongs to Bob.
The dotted box represents noisy channels, which makes the quantum channel to be mixed
state. The unitary operator $\tilde{U}$ makes $|\tilde{\psi}_W\rangle$ coincide with
$|\tilde{\psi}_{GHZ}\rangle$ expressed in Fig. 2}
\end{figure}


In this section we would like to repeat calculation of the previous section when 
$|\psi_{GHZ}\rangle$ is replaced by $|\psi_W\rangle$. In order to compute 
$F(\theta, \phi)$ we need a quantum circuit, which should be, of course, different 
from  Fig. 2. The quantum circuit for the quantum teleportation with $|\psi_W\rangle$
described in Fig. 3 is not simple like GHZ state. It cannot be represented by usual 
control-Not and Hardmard gates. In fact, we don't know how to express $\tilde{U}$-gate
described in Fig. 3 as combination of usual well-known gates such as control-Not, 
Hardmard, Pauli X, Y, Z and Toffoli gates. The $\tilde{U}$-gate is made to make 
$|\tilde{\psi}_W\rangle$ in Fig. 3 equal to $|\tilde{\psi}_{GHZ}\rangle$ in Fig 2. 
The explicit expression for $\tilde{U}$-gate is 
\begin{eqnarray}
\label{utilde}
\tilde{U} = \frac{1}{2} \left(            \begin{array}{cccccccc}
0 & 1 & 1 & 0 & \sqrt{2} & 0 & 0 & 0    \\
0 & 0 & 0 & 2 & 0 & 0 & 0 & 0           \\
0 & 0 & 0 & 0 & 0 & 0 & 0 & 2           \\
\sqrt{2} & 0 & 0 & 0 & 0 & 1 & 1 & 0     \\
0 & 1 & 1 & 0 & -\sqrt{2} & 0 & 0 & 0     \\
0 & \sqrt{2} & -\sqrt{2} & 0 & 0 & 0 & 0 & 0  \\
0 & 0 & 0 & 0 & 0 & \sqrt{2} & -\sqrt{2} & 0  \\
\sqrt{2} & 0 & 0 & 0 & 0 & -1 & -1 & 0
\end{array}              \right).
\end{eqnarray} 
In spite of, therefore, lack of knowledge on $\tilde{U}$-gate $\rho_{out}$, the 
density matrix for output state, can be derived by 
\begin{equation}
\label{output-w1}
\rho_{out} = \mbox{Tr}_{1,2,3} \left[U_W \rho_{in} \otimes \varepsilon (\rho_W)
U_W^{\dagger} \right]
\end{equation}
where the unitary operator $U_W$ can be read easily from Fig. 3 and 
$\varepsilon (\rho_W)$ is a density matrix constructed by 
$\rho_W\equiv |\psi_W\rangle \langle \psi_W|$ and noisy channels described by the 
dotted box in Fig. 3. 

Now we first consider $(L_{2,z}, L_{3,z}, L_{4,z})$ channel. In this case the 
master equation (\ref{lindbald}) with assuming, for simplicity, 
$\kappa_{2,z}=\kappa_{3,z}=\kappa_{4,z}=\kappa$ reduces to the simple first-order
differential equations, which gives 
\begin{eqnarray}
\label{qo-w1}
\varepsilon(\rho_W) = \frac{1}{4}
\left(                  \begin{array}{cccccccc}
0 & 0 & 0 & 0 & 0 & 0 & 0 & 0                 \\
0 & 2 & \sqrt{2} e^{-4 \kappa t} & 0 & \sqrt{2} e^{-4 \kappa t} & 0 & 0 & 0  \\
0 & \sqrt{2} e^{-4 \kappa t} & 1 & 0 & e^{-4 \kappa t} & 0 & 0 & 0           \\
0 & 0 & 0 & 0 & 0 & 0 & 0 & 0                 \\
0 & \sqrt{2} e^{-4 \kappa t} & e^{-4 \kappa t} & 0 & 1 & 0 & 0 & 0           \\
0 & 0 & 0 & 0 & 0 & 0 & 0 & 0                 \\
0 & 0 & 0 & 0 & 0 & 0 & 0 & 0                 \\
0 & 0 & 0 & 0 & 0 & 0 & 0 & 0                
\end{array}                     \right).
\end{eqnarray} 
Then Eq.(\ref{output-w1}) allows us to compute $\rho_{out}$ directly and 
Eq.(\ref{fidelity1}) gives
\begin{eqnarray}
\label{f-w1}
& &F(\theta, \phi) = 1 - \frac{1}{4} \left(1 - e^{-4 \kappa t} \right)
(1 + \sin^2 \theta)
\\   \nonumber
& &\bar{F} = \frac{1}{12} \left(7 + 5 e^{-4 \kappa t}\right).
\end{eqnarray}

Next we consider $(L_{2,x}, L_{3,x}, L_{4,x})$ noisy channel with 
$\kappa_{2,x}=\kappa_{3,x}=\kappa_{4,x}=\kappa$. In this case the master equation
reduces to $8$ diagonal coupled equations and $28$ off-diagonal coupled equations. The
diagonal equations imply $\sum_{i=0}^{3} \sigma_{ii} = 1/2 + e^{-2 \kappa t} / 4$ and
$\sum_{i=4}^{7} \sigma_{ii} = 1/2 - e^{-2 \kappa t} / 4$, where 
$\sigma_{ij} = \varepsilon_{ij} (\rho_W)$ with $i,j = 0, \cdots, 7$. Using these two
constraints one can compute all diagonal components, which are 
\begin{eqnarray}
\label{sw1}
& &\sigma_{00} = \frac{1}{8} \left(1 + e^{-2 \kappa t} - e^{-4 \kappa t} 
- e^{-6 \kappa t} \right)
\\   \nonumber
& &\sigma_{11} = \frac{1}{8} \left(1 + e^{-2 \kappa t} + e^{-4 \kappa t} 
+ e^{-6 \kappa t} \right)
\\   \nonumber
& &\sigma_{22} = \sigma_{44} = \frac{1}{8} (1 + e^{-6 \kappa t} )
\\   \nonumber
& &\sigma_{33} = \sigma_{55} = \frac{1}{8} (1 - e^{-6 \kappa t} )
\\   \nonumber
& &\sigma_{66} = \frac{1}{8} \left(1 - e^{-2 \kappa t} + e^{-4 \kappa t} 
- e^{-6 \kappa t} \right)
\\   \nonumber
& &\sigma_{77} = \frac{1}{8} \left(1 - e^{-2 \kappa t} - e^{-4 \kappa t} 
+ e^{-6 \kappa t} \right).
\end{eqnarray}                                                              

The equations for the off-diagonal components are more complicated. However, these 
equations consist of $7$ groups, each of which are $4$ closed coupled equations. This 
fact allows us to compute all components analytically, whose explicit expressions are 
\begin{eqnarray}
\label{sw2}
& &\sigma_{03} = \sigma_{05} = \sqrt{2} \sigma_{06} = \frac{\sqrt{2}}{16}
\left(1 + e^{-2 \kappa t} - e^{-4 \kappa t} - e^{-6 \kappa t} \right)
\\  \nonumber
& &\sigma_{12} = \sigma_{14} = \sqrt{2} \sigma_{24} = \frac{\sqrt{2}}{16}
\left(1 + e^{-2 \kappa t} + e^{-4 \kappa t} + e^{-6 \kappa t} \right)
\\  \nonumber
& &\sigma_{27} = \sigma_{47} = \sqrt{2} \sigma_{17} = \frac{\sqrt{2}}{16}
\left(1 - e^{-2 \kappa t} - e^{-4 \kappa t} + e^{-6 \kappa t} \right)
\\  \nonumber
& &\sigma_{36} = \sigma_{56} = \sqrt{2} \sigma_{35} = \frac{\sqrt{2}}{16}
\left(1 - e^{-2 \kappa t} + e^{-4 \kappa t} - e^{-6 \kappa t} \right)
\end{eqnarray}
with $\sigma_{ij} = \sigma_{ji}$.
Defining                                                               
\begin{eqnarray}
\label{def-w1}
& &\alpha_1 = 1 + e^{-2 \kappa t} + e^{-4 \kappa t} + e^{-6 \kappa t}  \\  \nonumber
& &\alpha_2 = 1 + e^{-2 \kappa t} - e^{-4 \kappa t} - e^{-6 \kappa t}  \\  \nonumber
& &\alpha_3 = 1 - e^{-2 \kappa t} - e^{-4 \kappa t} + e^{-6 \kappa t}  \\  \nonumber
& &\alpha_4 = 1 - e^{-2 \kappa t} + e^{-4 \kappa t} - e^{-6 \kappa t}  \\  \nonumber
& &\beta_{\pm} = 1 \pm e^{-6 \kappa t},
\end{eqnarray}
one can express $\varepsilon(\rho_W)$ as following:
\begin{eqnarray}
\label{qo-w2}
\varepsilon(\rho_W) = \frac{1}{16}
\left(           \begin{array}{cccccccc}
2 \alpha_2 & 0 & 0 & \sqrt{2} \alpha_2 & 0 & \sqrt{2} \alpha_2 & \alpha_2 & 0  \\
0 & 2 \alpha_1 & \sqrt{2} \alpha_1 & 0 & \sqrt{2} \alpha_1 & 0 & 0 & \alpha_3  \\
0 & \sqrt{2} \alpha_1 & 2 \beta_+ & 0 & \alpha_1 & 0 & 0 & \sqrt{2} \alpha_3  \\
\sqrt{2} \alpha_2 & 0 & 0 & 2 \beta_- & 0 & \alpha_4 & \sqrt{2} \alpha_4 & 0   \\
0 & \sqrt{2} \alpha_1 & \alpha_1 & 0 & 2 \beta_+ & 0 & 0 & \sqrt{2} \alpha_3   \\
\sqrt{2} \alpha_2 & 0 & 0 & \alpha_4 & 0 & 2 \beta_- & \sqrt{2} \alpha_4 & 0   \\
\alpha_2 & 0 & 0 & \sqrt{2} \alpha_4 & 0 & \sqrt{2} \alpha_4 & 2 \alpha_4 & 0  \\
0 & \alpha_3 & \sqrt{2} \alpha_3 & 0 & \sqrt{2} \alpha_3 & 0 & 0 & 2 \alpha_3
\end{array}
\right).
\end{eqnarray}
Inserting Eq.(\ref{qo-w2}) into (\ref{output-w1}), one can compute $\rho_{out}$
directly. Thus using $\rho_{out}$ and Eq.(\ref{fidelity1}), one can compute 
$F(\theta, \phi)$ and $\bar{F}$ whose expressions are 
\begin{eqnarray}
\label{f-w2}
& &F(\theta, \phi) = \frac{1}{8} \bigg[ (4 + 2 \sin^2 \theta \cos^2 \phi) + 
e^{-2 \kappa t} (\cos^2 \theta + 2 \sin^2 \theta \cos^2 \phi)
\\   \nonumber
& &\hspace{2.0cm}
+ e^{-4 \kappa t} (2 \sin^2 \theta \sin^2 \phi) + 
+ e^{-6 \kappa t} (3 \cos^2 \theta + 2 \sin^2 \theta \sin^2 \phi)
\bigg]
\\   \nonumber
& &\bar{F} = \frac{1}{24} \left(14 + 3 e^{-2 \kappa t} + 2 e^{-4 \kappa t} + 5 
e^{-6 \kappa t} \right).
\end{eqnarray}

For $(L_{2,y},L_{3,y},L_{4,y})$ noisy channel similar calculation shows that 
$\varepsilon(\rho_w)$ reduces to
\begin{eqnarray}
\label{qo-w3}
\varepsilon(\rho_W) = \frac{1}{16}
\left(           \begin{array}{cccccccc}
2 \alpha_2 & 0 & 0 & -\sqrt{2} \alpha_2 & 0 & -\sqrt{2} \alpha_2 & -\alpha_2 & 0  \\
0 & 2 \alpha_1 & \sqrt{2} \alpha_1 & 0 & \sqrt{2} \alpha_1 & 0 & 0 & -\alpha_3  \\
0 & \sqrt{2} \alpha_1 & 2 \beta_+ & 0 & \alpha_1 & 0 & 0 & -\sqrt{2} \alpha_3  \\
-\sqrt{2} \alpha_2 & 0 & 0 & 2 \beta_- & 0 & \alpha_4 & \sqrt{2} \alpha_4 & 0   \\
0 & \sqrt{2} \alpha_1 & \alpha_1 & 0 & 2 \beta_+ & 0 & 0 & -\sqrt{2} \alpha_3   \\
-\sqrt{2} \alpha_2 & 0 & 0 & \alpha_4 & 0 & 2 \beta_- & \sqrt{2} \alpha_4 & 0   \\
-\alpha_2 & 0 & 0 & \sqrt{2} \alpha_4 & 0 & \sqrt{2} \alpha_4 & 2 \alpha_4 & 0  \\
0 & -\alpha_3 & -\sqrt{2} \alpha_3 & 0 & -\sqrt{2} \alpha_3 & 0 & 0 & 2 \alpha_3
\end{array}
\right)
\end{eqnarray}
and, as a result, $F(\theta, \phi)$ and $\bar{F}$ reduce to
\begin{eqnarray}
\label{f-w3}
& &F(\theta, \phi) = \frac{1}{8} \bigg[ (4 + 2 \sin^2 \theta \sin^2 \phi) + 
e^{-2 \kappa t} (\cos^2 \theta + 2 \sin^2 \theta \sin^2 \phi)
\\   \nonumber
& &\hspace{2.0cm}
+ e^{-4 \kappa t} (2 \sin^2 \theta \cos^2 \phi) +
+ e^{-6 \kappa t} (3 \cos^2 \theta + 2 \sin^2 \theta \cos^2 \phi)
\bigg]
\\   \nonumber
& &\bar{F} = \frac{1}{24} \left(14 + 3 e^{-2 \kappa t} + 2 e^{-4 \kappa t} + 5 
e^{-6 \kappa t} \right).
\end{eqnarray}
It is interesting to note that $\bar{F}$ for $(L_{2,x},L_{3,x},L_{4,x})$ noisy
channel is same with $\bar{F}$ for $(L_{2,y},L_{3,y},L_{4,y})$ noisy channel.

Finally for isotropic noisy channel $\varepsilon(\rho_W)$ becomes
\begin{eqnarray}
\label{qo-w4}
\varepsilon(\rho_W) = \frac{1}{8}
\left(           \begin{array}{cccccccc}
\tilde{\alpha}_2 & 0 & 0 & 0 & 0 & 0 & 0 & 0  \\
0 & \tilde{\alpha}_1 & \sqrt{2} \tilde{\gamma}_+ & 0 & \sqrt{2} \tilde{\gamma}_+ 
& 0 & 0 & 0  \\
0 & \sqrt{2} \tilde{\gamma}_+ & \tilde{\beta}_+ & 0 & \tilde{\gamma}_+ & 0 & 0 & 0  \\
0 & 0 & 0 & \tilde{\beta}_- & 0 & \tilde{\gamma}_- & \sqrt{2} \tilde{\gamma}_- & 0   \\
0 & \sqrt{2} \tilde{\gamma}_+ & \tilde{\gamma}_+ & 0 & \tilde{\beta}_+ & 0 & 0 & 0   \\
0 & 0 & 0 & \tilde{\gamma}_- & 0 & \tilde{\beta}_- & \sqrt{2} \tilde{\gamma}_- & 0   \\
0 & 0 & 0 & \sqrt{2} \tilde{\gamma}_- & 0 & \sqrt{2} \tilde{\gamma}_- 
& \tilde{\alpha}_4 & 0  \\
0 & 0 & 0 & 0 & 0 & 0 & 0 & \tilde{\alpha}_3
\end{array}
\right)
\end{eqnarray}
where
\begin{eqnarray}
\label{def-w2}
& &\tilde{\alpha}_1 = 1 + e^{-4 \kappa t} + e^{-8 \kappa t} + e^{-12 \kappa t}
\\   \nonumber
& &\tilde{\alpha}_2 = 1 + e^{-4 \kappa t} - e^{-8 \kappa t} - e^{-12 \kappa t}
\\   \nonumber
& &\tilde{\alpha}_3 = 1 - e^{-4 \kappa t} - e^{-8 \kappa t} + e^{-12 \kappa t}
\\   \nonumber
& &\tilde{\alpha}_4 = 1 - e^{-4 \kappa t} + e^{-8 \kappa t} - e^{-12 \kappa t}
\\   \nonumber
& &\tilde{\beta}_{\pm} = 1 \pm e^{-12 \kappa t}              \\   \nonumber
& &\tilde{\gamma}_{\pm} = e^{-8 \kappa t} \pm e^{-12 \kappa t}.
\end{eqnarray}
Thus one can compute $F(\theta, \phi)$ and $\bar{F}$ for this noisy channel, which
are
\begin{eqnarray}
\label{f-w4}
& &F(\theta, \phi) = \frac{1}{4} \left[ 2 + e^{-8 \kappa t} \sin^2 \theta + 
e^{-12 \kappa t} (1 + \cos^2 \theta)  \right]
\\  \nonumber
& &\bar{F} = \frac{1}{6} \left(3 + e^{-8 \kappa t} + 2 e^{-12 \kappa t} \right).
\end{eqnarray}
The measures $F(\theta, \phi)$ and $\bar{F}$ for the various noisy channels are 
summarized in Table I with those for GHZ state. In the next section we will compare
$F(\theta, \phi)$ and $\bar{F}$ for GHZ state with those for W state.

\newpage

\section{GHZ versus W}

\begin{center}
\begin{tabular}{c|c|c|c}  \hline
{} & noise & GHZ & W \\ \hline \hline
{} & ($(L_{2x}$,$L_{3x}$,$L_{4x}$) & $\frac{1}{2} \bigg[
(1 + \sin^2 \theta \cos^2 \phi$) & 
$\frac{1}{8} \bigg[(4 + 2 \sin^2 \theta \cos^2 \phi)$ \\
{} & {} & $+ e^{-4 \kappa t} (1 - \sin^2 \theta \cos^2 \phi) \bigg]$ & $ 
+ e^{-2 \kappa t} (\cos^2 \theta 
+ 2 \sin^2 \theta \cos^2 \phi)$  
\\  
{} & {} & {} & 
$+ e^{-4 \kappa t} (2 \sin^2 \theta \sin^2 \phi) $  \\
{} & {} & {} & $+
e^{-6 \kappa t} (3 \cos^2 \theta + 2 \sin^2 \theta \sin^2 \phi)\bigg] $ \\ \cline{2-4}
$F(\theta,\phi)$ & ($(L_{2y}$,$L_{3y}$,$L_{4y}$) & $\frac{1}{2} \bigg[1 + 
\sin^2 \theta \sin^2 \phi e^{-2 \kappa t}$   & $\frac{1}{8} \bigg[ 
(4 + 2 \sin^2 \theta \sin^2 \phi)$    \\ 
{} & {} & $+ \cos^2 \theta e^{-4 \kappa t}$ & $+ e^{-2 \kappa t} (\cos^2 \theta + 2 
\sin^2 \theta \sin^2 \phi)$                                           \\
{} & {} & $+ \sin^2 \theta \cos^2 \phi e^{-6 \kappa t} \bigg]$ & 
$+ e^{-4 \kappa t} (2 \sin^2 \theta \cos^2 \phi)$                      \\
{} & {} & {} & $+ e^{-6 \kappa t} (3 \cos^2 \theta + 2 \sin^2 \theta \cos^2 \phi) 
\bigg]$
\\ \cline{2-4}
{} & ($(L_{2z}$,$L_{3z}$,$L_{4z}$) & $1 - \frac{1}{2} (1 - e^{-6 \kappa t}) 
\sin^2 \theta$  &   $1 - \frac{1}{4} (1 - e^{-4 \kappa t}) (1 + \sin^2 \theta)$ 
\\ \cline{2-4}
{} & isotropy & $\frac{1}{2} (1 + \cos^2 \theta e^{-8 \kappa t} + \sin^2 \theta
e^{-12 \kappa t} )$ & 
$\frac{1}{4} [2 + \sin^2 \theta e^{-8 \kappa t} + (1 + \cos^2 \theta) 
e^{-12 \kappa t} ]$
\\  \hline \hline
{} & ($(L_{2x}$,$L_{3x}$,$L_{4x}$) & $\frac{2}{3} + \frac{1}{3} e^{-4 \kappa t}$ &
$\frac{1}{24} (14 + 3 e^{-2 \kappa t} + 2 e^{-4 \kappa t} + 5 e^{-6 \kappa t})$
\\ \cline{2-4}
$\bar{F}$ & ($(L_{2y}$,$L_{3y}$,$L_{4y}$) &   $\frac{1}{6} (3 + e^{-2 \kappa t} +
e^{-4 \kappa t} + e^{-6 \kappa t})$    & $\frac{1}{24} (14 + 3 e^{-2 \kappa t} + 
2 e^{-4 \kappa t} + 5 e^{-6 \kappa t})$ 
\\ \cline{2-4}
{} & ($(L_{2z}$,$L_{3z}$,$L_{4z}$) & $\frac{2}{3} + \frac{1}{3} e^{-6 \kappa t}$ &
$\frac{1}{12} (7 + 5 e^{-4 \kappa t})$
\\  \cline{2-4}
{} & isotropy & $\frac{1}{6} (3 + e^{-8 \kappa t} + 2 e^{-12 \kappa t} )$ &
$\frac{1}{6} (3 + e^{-8 \kappa t} + 2 e^{-12 \kappa t})$
\\  \hline
\end{tabular}

\vspace{0.1cm}
Table I: Summary of $F(\theta, \phi)$ and $\bar{F}$ in various noisy channels.
\end{center}
\vspace{0.5cm}

The quantities $F(\theta, \phi)$ and $\bar{F}$ for various noisy channels are 
summarized at Table I when GHZ and W states are prepared for the quantum teleportation.
The most interesting feature in Table I is the fact that $\bar{F}$ for GHZ is exactly 
same with that for W in the isotropic channel. Since the isotropic noisy channel can
be regarded roughly as a sum of $(L_{2,x}, L_{3,x}, L_{4,x})$, 
$(L_{2,y}, L_{3,y}, L_{4,y})$, and $(L_{2,z}, L_{3,z}, L_{4,z})$ noisy channels, 
this fact indicates that which state between GHZ and W does not lose information is
noise-dependent.

\begin{figure}[ht!]
\begin{center}
\includegraphics[height=6.3cm]{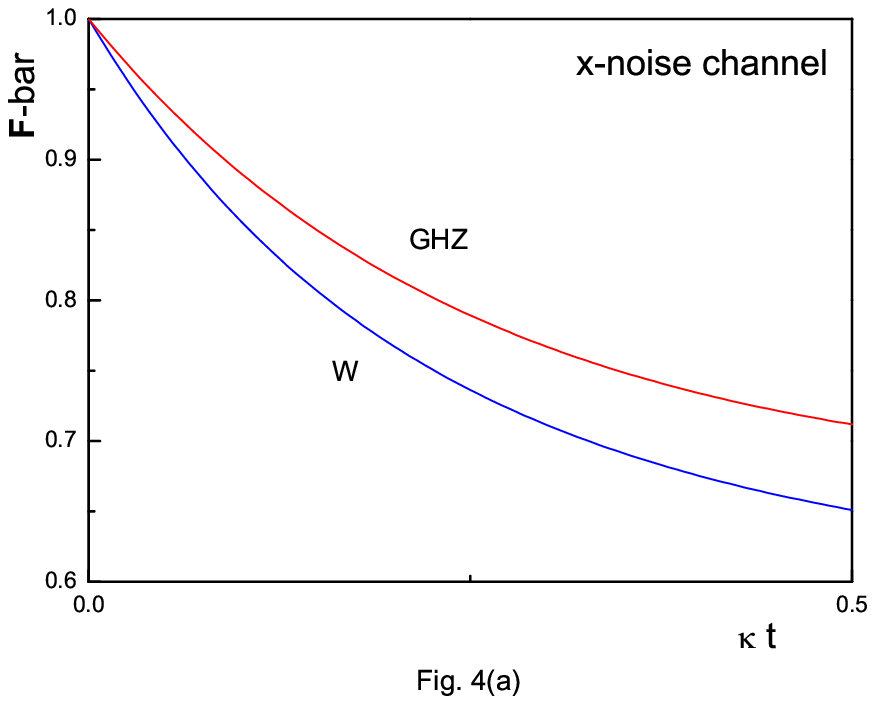}
\includegraphics[height=6.3cm]{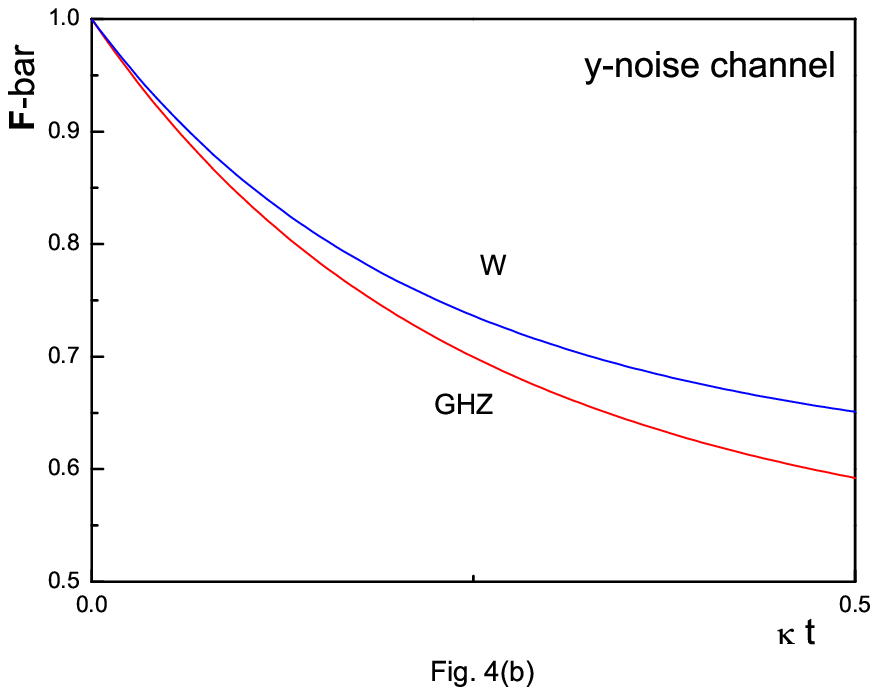}
\includegraphics[height=6.3cm]{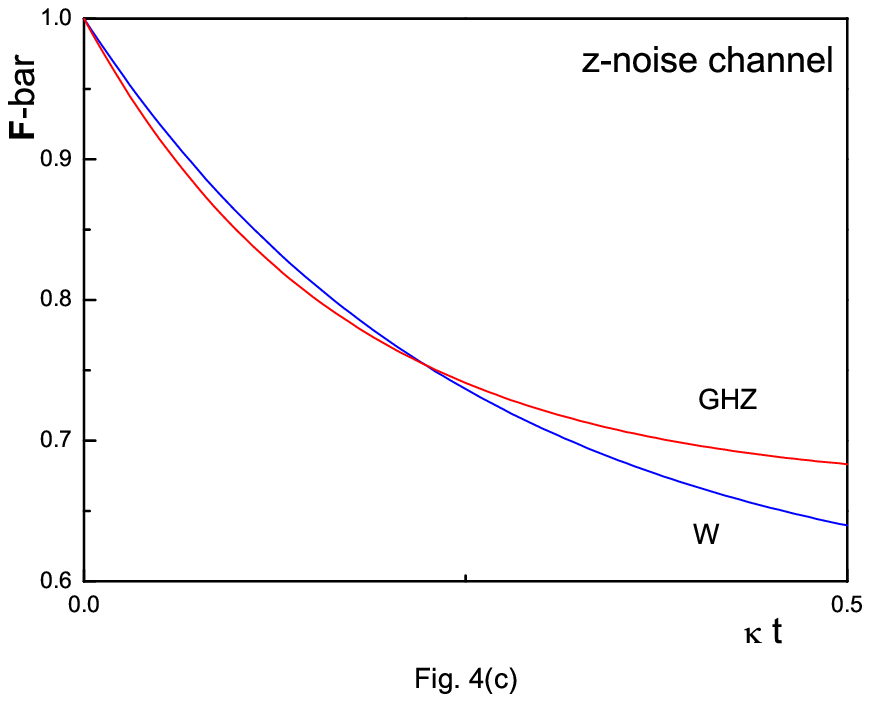}
\caption[fig4]{The plot of $\kappa t$-dependence of $\bar{F}$ for 
($L_{2,x}$,$L_{3,x}$,$L_{4,x}$) (Fig. 4a), ($L_{2,y}$,$L_{3,y}$,$L_{4,y}$) (Fig. 4b), and 
($L_{2,z}$,$L_{3,z}$,$L_{4,z}$) (Fig. 4c) noisy channels. Fig. 4a shows that $\bar{F}$
for GHZ state is always larger than that for W state, which implies that GHZ state does
lose less quantum information than W state in ($L_{2,x}$,$L_{3,x}$,$L_{4,x}$) noisy 
channel. Fig. 4b shows, however, that the situation is completely reversed in 
($L_{2,y}$,$L_{3,y}$,$L_{4,y}$) noisy channel. In ($L_{2,z}$,$L_{3,z}$,$L_{4,z}$) 
noisy channel Fig. 4c indicates that W state is more robust when $\kappa t < 0.223$
while GHZ state becomes more robust when $\kappa t > 0.223$.}
\end{center}
\end{figure}

In order to show this fact explicitly we plot the $\kappa t$-dependence of $\bar{F}$
for $(L_{2,x}, L_{3,x}, L_{4,x})$ (Fig. 4a), $(L_{2,y}, L_{3,y}, L_{4,y})$ (Fig. 4b), 
and $(L_{2,z}, L_{3,z}, L_{4,z})$ (Fig. 4c) noisy channels. Fig. 4 shows that $\bar{F}$ 
for $|\psi_{GHZ}\rangle$ is always larger than that for $|\psi_W\rangle$ in 
$(L_{2,x}, L_{3,x}, L_{4,x})$ noisy channel. This means that $|\psi_{GHZ}\rangle$ does 
lose less quantum information compared to $|\psi_W\rangle$ in this noisy channel. 
However, the situation is changed in $(L_{2,y}, L_{3,y}, L_{4,y})$ noisy channel. In 
this case $\bar{F}$ for $|\psi_W\rangle$ is always larger than that for 
$|\psi_{GHZ}\rangle$. This means that $|\psi_W\rangle$ is more robust than 
$|\psi_{GHZ}\rangle$ in this noisy channel. In $(L_{2,z}, L_{3,z}, L_{4,z})$ noisy 
channel the situation is more delicate. In this channel $\bar{F}$  for $|\psi_W\rangle$
is larger than that for $|\psi_{GHZ}\rangle$ when $\kappa t \leq 0.223$. If, however, 
$\kappa t \geq 0.223$, $\bar{F}$  for $|\psi_{GHZ}\rangle$ becomes larger than that
for $|\psi_W\rangle$. Summing over all those phenomena seems to make same $\bar{F}$
for $|\psi_{GHZ}\rangle$ and $|\psi_W\rangle$ in the isotropic channel.

However, we should note that the result of Fig. 4 is dependent on the choice of the 
basis. To show this explicitly let us consider an unitary operator
$U = (\sigma_x + \sigma_y) / \sqrt{2}$, which yields $U \sigma_x U^{\dagger} = \sigma_y$ and
$U \sigma_y U^{\dagger} = \sigma_x$. Now, let us consider the noisy teleportation when 
quantum channels are $|\psi^{\prime}_{GHZ}\rangle = U \otimes U \otimes U |\psi_{GHZ}\rangle$
and $|\psi^{\prime}_{W}\rangle = U \otimes U \otimes U |\psi_{W}\rangle$ respectively.
Then it is obvious that Fig. 4a and Fig. 4b would be interchanged with each other if one
computes the average fidelity. This indicates that Fig. 4 is dependent on the choice of the
basis states. 

\begin{figure}[ht!]
\begin{center}
\caption[fig5]{The plot of ($\theta$, $\phi$)-dependence of $F(\theta, \phi)$ for 
$(L_{2,x}, L_{3,x}, L_{4,x})$ (Fig. 5a), $(L_{2,y}, L_{3,y}, L_{4,y})$ (Fig. 5b),
$(L_{2,z}, L_{3,z}, L_{4,z})$ (Fig. 5c), and isotropic (Fig. 5d) noisy channels. The 
transparent and opaque surfaces correspond to GHZ and W states respectively. All 
figures are consistent with $\bar{F}$ given in Table I.}
\end{center}
\end{figure}

Another interesting point in Table I is the fact that $\bar{F}$ for GHZ state decays to 
$2/3$ in $(L_{2,x}, L_{3,x}, L_{4,x})$ and $(L_{2,z}, L_{3,z}, L_{4,z})$ noisy channels. 
The number $\bar{F} = 2/3$ corresponds to the average fidelity obtained only by the 
classical communication\cite{mass94}. However, in $(L_{2,y}, L_{3,y}, L_{4,y})$ noisy 
channel $\bar{F}$ for GHZ state decays to $1/2$, which corresponds to no-communication
between Alice and Bob. When quantum channel is subject to noise in one direction, 
the average fidelity for W state always decays to $7/12$, which is slightly smaller 
than $2/3$. In isotropic noisy channel $\bar{F}$ for both GHZ and W states decays to 
$1/2$ when $\kappa t \rightarrow \infty$ like two-qubit EPR quantum channel\cite{oh02}.

Fig. 5 is plot of $\theta$- and $\phi$-dependence of $F(\theta, \phi)$ for 
$(L_{2,x}, L_{3,x}, L_{4,x})$ (Fig. 5a), $(L_{2,y}, L_{3,y}, L_{4,y})$ (Fig. 5b), 
$(L_{2,z}, L_{3,z}, L_{4,z})$ (Fig. 5c) and isotropic (Fig. 5d) noisy channels when
$\kappa t$ is fixed to $0.5$. The transparent and opaque surfaces correspond to 
GHZ and W states respectively. Fig. 5a indicates that in $(L_{2,x}, L_{3,x}, L_{4,x})$
noisy channel $F(\theta, \phi)$ for GHZ state is larger than that for W state
in entire range of $\theta$ and $\phi$. Fig. 5b shows that in 
$(L_{2,y}, L_{3,y}, L_{4,y})$ noisy channel $F(\theta, \phi)$ for W state is 
larger in almost range of $\theta$ and $\phi$ except small boundary region. This is 
consistent with the fact that $\bar{F}$ for W state is larger than that for GHZ
state in this noisy channel. Fig. 5c and Fig. 5d shows that in 
$(L_{2,z}, L_{3,z}, L_{4,z})$ and isotropic noisy channels $F(\theta, \phi)$ for 
GHZ state is generally larger than that for W state in small $\theta$ region
(approximately $0 \leq \theta < 1$) and large $\theta$ region (approximately 
$2 < \theta \leq \pi$) while in the middle $\theta$ region (approximately
$1 < \theta < 2$) $F(\theta, \phi)$ for W state is larger.

\section{conclusion}

\begin{figure}[ht!]
\begin{center}
\includegraphics[height=6.5cm]{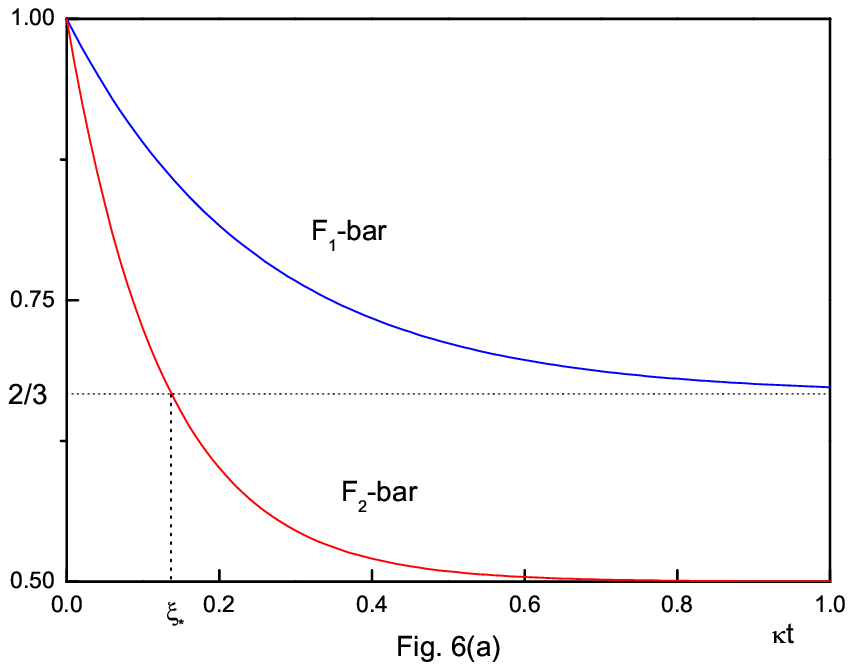}
\includegraphics[height=6.5cm]{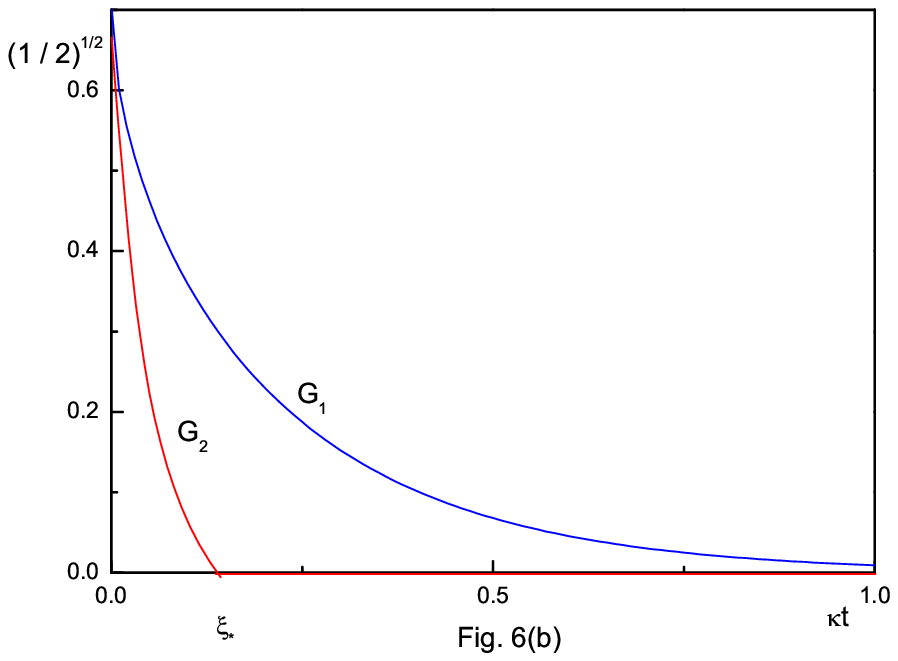}
\caption[fig6]{Conjecture of Relation between $\bar{F}$ and mixed states Groverian
measure. Since $\bar{F}_2$ becomes smaller than $2/3$ when 
$\kappa t \geq \xi_* = - \ln (\sqrt{2} - 1) / 2$, the corresponding Groverian measure $G_2$ is 
expected to vanish in the same region.}
\end{center}
\end{figure}

In this paper we consider the quantum teleportation with GHZ and W states respectively, 
when the noisy channels make the quantum channels to be mixed states. The issue of 
robustness between GHZ and W, i.e. which state does lose less quantum 
information, in the noisy channels is completely dependent on the type of noisy.
If, for example, the noisy channel is $(L_{2,x}, L_{3,x}, L_{4,x})$-type, GHZ state is 
always robust compared to W state while the reverse situation occurs in 
$(L_{2,y}, L_{3,y}, L_{4,y})$ noisy channel. In $(L_{2,z}, L_{3,z}, L_{4,z})$ noisy
channel W state does lose less information than GHZ state when $\kappa t$ is 
comparatively small. If, however, $\kappa t \geq 0.223$, GHZ state becomes more robust
in this noisy channel. 

Since the decoherence mechanism in each qubit is obviously independent, one can 
explore the different noisy channels for each qubit in the given quantum channel such as
$(L_{2,x}, L_{3,y}, L_{4,z})$ noisy channel. In this sense the noisy channels discussed 
in this paper can be said to be oversimplified. The reason that we consider only the noisy
channels with same axis in this paper can be summarized as following. First, 
the main purpose of this paper is to show explicitly that the robustness between GHZ and 
W states in the noisy teleportation is dependent on the noisy types. Thus, as shown in Fig. 4
it is sufficient to introduce the same-axis noisy channels.
Another reason is that we would like to explore the cases of high fidelity because the 
quantum channels become useless if $\bar{F}$ is comparatively small. We conjecture that 
$\bar{F}$ with same-axis noisy channels are in general  larger than $\bar{F}$ with 
different-axis noisy channels. For example, let us consider the teleportation with 
EPR state depicted in Fig. 1. When the quantum channel is subject to $(L_{2,x}, L_{3,x})$
or $(L_{2,z}, L_{3,z})$ noisy channels, the average fidelity $\bar{F}$ is always 
\begin{equation}
\label{boso1}
\bar{F}_1 = \frac{2}{3} + \frac{1}{3} e^{-4 \kappa t}.
\end{equation}
If, however, the quantum channel is subject to $(L_{2,x}, L_{3,z})$ or $(L_{2,z}, L_{3,x})$
noisy channels, direct calculation shows that the average fidelity reduces to 
\begin{equation}
\label{cc1}
\bar{F}_D = \frac{1}{6} \left( 3 + 2 e^{-2 \kappa t} + e^{-4 \kappa t} \right),
\end{equation}
which is smaller than $\bar{F}_1$ in full range of $\kappa t$. This supports our conjecture
although detailed calculation is needed for the complete proof.

Probably one may be able to increase $F(\theta, \phi)$ and $\bar{F}$ summarized in 
Table I via the purification of noisy channels discussed in Ref.\cite{bennett95,cheong07}.
To explore this issue, of course, we need another detailed calculation, which is beyond
the scope of present paper.

It is of interest to extend our papers to examine the fidelity measures 
$F(\theta, \phi)$ and $\bar{F}$ when other types of noisy channels such as amplitude 
damping or depolarizing channels are introduced. It is also equally interest to 
examine the same noisy channels in other places such as noisy channels during Bell's 
measurement or the unitary operation. 

The most important point we would like to explore 
in the future is to understand the physical reason why and how the robustness of GHZ 
and W states is dependent on the noisy-types. In our opinion the most nice approach to
understand the physical reason is to investigate the entanglement of the mixed states
$\varepsilon(\rho_{GHZ})$ and $\varepsilon(\rho_W)$. For example, let us consider the 
quantum teleportation through the noisy channels with EPR state for brevity, which is 
fully discussed in Ref.\cite{oh02}. In this case when the quantum channel is subject to 
$(L_{2,x}, L_{3,x})$, $(L_{2,y}, L_{3,y})$ or $(L_{2,z}, L_{3,z})$ noisy channels, the 
average fidelity $\bar{F}$ is always same with Eq.(\ref{boso1}),
while the isotropic noisy channel gives
\begin{equation}
\label{boso2}
\bar{F}_2 = \frac{1}{2} + \frac{1}{2} e^{-8 \kappa t}.
\end{equation}
Then we think that an appropriate entanglement measure should have following properties.
The measure for the mixed state $\varepsilon_1(\rho)$ generated by $(L_{2,x}, L_{3,x})$, 
$(L_{2,y}, L_{3,y})$ and $(L_{2,z}, L_{3,z})$ noisy channels should decay to zero at 
$\kappa t \rightarrow \infty$ because $\bar{F} = 2/3$ implies that the mixed states
do not play any role as quantum channels. By same reason the measure for the 
mixed state $\varepsilon_2(\rho)$ generated by the isotropic noisy channel should vanish
at $\kappa t \geq (1/8) \ln 3$. 

If we take a Groverian entanglement measure $G(\rho)$\cite{biham01-1,shapira06} as an
entanglement measure, there is another constraint $G(\rho) = 1/\sqrt{2}$ at $\kappa t = 0$
because the Groverian measure for the pure EPR state is $1/\sqrt{2}$. As a result, we 
can conjecture that the Groverian measure $G_1$ and $G_2$ for $\varepsilon_1(\rho)$ and 
$\varepsilon_2(\rho)$ may exhibit as Fig. 6. We would like to show whether or not our 
conjecture is correct. In addition we would like to extend our conjecture to the 
quantum teleportation through noisy channels with GHZ and W states discussed in this 
paper.

{\bf Acknowledgement}: 
This work was supported by the Kyungnam University
Foundation Grant, 2008.

\newpage

\begin{appendix}{\centerline{\bf Appendix }}

\setcounter{equation}{0}
\renewcommand{\theequation}{A.\arabic{equation}}

Let, for simplicity, $\sigma_{ij}^x$ and $\sigma_{ij}^y$ be the density matrices for 
$(L_{2x},L_{3x},L_{4x})$ and $(L_{2y},L_{3y},L_{4y})$ noises respectively. Then the 
master equation (\ref{lindbald}) makes the off-diagonal components of $\sigma_{ij}^x$ and 
$\sigma_{ij}^y$ satisfy the following coupled equations:
\begin{eqnarray}
\label{app1}
& & \frac{d \sigma_{07}^x}{dt} = -\kappa (3 \sigma_{07}^x  - \sigma_{16}^x 
                                            - \sigma_{25}^x - \sigma_{43}^x)
\\   \nonumber
& & \frac{d \sigma_{16}^x}{dt} = -\kappa (3 \sigma_{16}^x - \sigma_{07}^x 
                                           - \sigma_{34}^x - \sigma_{52}^x)
\\   \nonumber
& &\frac{d \sigma_{25}^x}{dt} = -\kappa (3 \sigma_{25}^x - \sigma_{07}^x 
                                           - \sigma_{34}^x - \sigma_{61}^x)
\\   \nonumber
& &\frac{d \sigma_{34}^x}{dt} = -\kappa (3 \sigma_{34}^x - \sigma_{16}^x 
                                           - \sigma_{25}^x - \sigma_{70}^x) 
\end{eqnarray}
and
\begin{eqnarray}
\label{app2}
& & \frac{d \sigma_{07}^y}{dt} = -\kappa (3 \sigma_{07}^y + \sigma_{16}^y 
                                            + \sigma_{25}^y + \sigma_{43}^y)
\\   \nonumber
& & \frac{d \sigma_{16}^y}{dt} = -\kappa (3 \sigma_{16}^y + \sigma_{07}^y 
                                            + \sigma_{34}^y + \sigma_{52}^y)
\\   \nonumber
& &\frac{d \sigma_{25}^y}{dt} = -\kappa (3 \sigma_{25}^y + \sigma_{07}^y 
                                           + \sigma_{34}^y + \sigma_{61}^y)
\\   \nonumber
& &\frac{d \sigma_{34}^y}{dt} = -\kappa (3 \sigma_{34}^y + \sigma_{16}^y 
                                           + \sigma_{25}^y + \sigma_{70}^y)
\end{eqnarray}
and their complex conjugates. Then it is easy to show that 
$\sigma_{07}^x=\sigma_{70}^x=\alpha_+ / 8$, $\sigma_{16}^x=\sigma_{61}^x
         =\sigma_{25}^x=\sigma_{52}^x=\sigma_{34}^x=\sigma_{43}^x=\alpha_-/8$,
$\sigma_{07}^y=\sigma_{70}^y=\beta_1/8$, and $\sigma_{16}^y=\sigma_{61}^y
         =\sigma_{25}^y=\sigma_{52}^y=\sigma_{34}^y=\sigma_{43}^y= -\beta_2/8$
satisfy Eq.(\ref{app1}) and Eq.(\ref{app2}). Also these solutions satisfy the boundary
condition $\sigma_{ij}^x = \sigma_{ij}^y = \rho_{GHZ}$ at $\kappa t = 0$.

If we ignore the boundary condition, many different solutions for $\sigma_{ij}^y$ can be 
obtained from $\sigma_{ij}^x$. For example, $\sigma_{07}^y=- \sigma_{70}^y=i \alpha_+$ and 
$\sigma_{16}^y=-\sigma_{61}^y=\sigma_{25}^y=-\sigma_{52}^y=\sigma_{43}^y
               =-\sigma_{34}^y=-i \alpha_-$ are also solutions of 
Eq.(\ref{app2}). These are the solutions derived from 
$(u\otimes u \otimes u) \sigma_{ij}^x (u\otimes u \otimes u)^{\dagger}$ when
\begin{eqnarray}
\label{app3}
u = \frac{1}{\sqrt{2}} \left(         \begin{array}{cc}
0   &    1 - i    \\
1 + i  &   0
\end{array}               \right).
\end{eqnarray}
Even if these are solutions of Eq.(\ref{app2}), they do not satisfy the proper
boundary condition.
\end{appendix}

\end{document}